\documentclass{article}
\usepackage[final]{graphics}
\usepackage{amsfonts,amsbsy}

\font\tenmsa=msam10
\font\sevenmsa=msam7
\font\fivemsa=msam5
\font\tenmsb=msbm10
\font\sevenmsb=msbm7
\font\fivemsb=msbm5
\newfam\msafam
\newfam\msbfam
\textfont\msafam=\tenmsa  \scriptfont\msafam=\sevenmsa
\scriptscriptfont\msafam=\fivemsa
\textfont\msbfam=\tenmsb  \scriptfont\msbfam=\sevenmsb
\scriptscriptfont\msbfam=\fivemsb

\catcode`\@=11

\newcount\@tempcntc
\def\@citex[#1]#2{\if@filesw\immediate\write\@auxout{\string\citation{#2}}\fi
  \@tempcnta\z@\@tempcntb\m@ne\def\@citea{}\@cite{%
        \@for\@citeb:=#2\do%
    {\@ifundefined{b@\@citeb}%
        {\@citeo\@tempcntb\m@ne\@citea%
                \def\@citea{,\penalty\@m\ }{\bf ?}\@warning%
                {Citation `\@citeb' on page \thepage \space undefined}}%
        {\setbox\z@\hbox{\global\@tempcntc0\csname b@\@citeb\endcsname\relax}
     \ifnum\@tempcntc=\z@ \@citeo\@tempcntb\m@ne%
       \@citea\def\@citea{,\penalty\@m}%
       \hbox{\csname b@\@citeb\endcsname}%
     \else%
      \advance\@tempcntb\@ne%
      \ifnum\@tempcntb=\@tempcntc%
      \else\advance\@tempcntb\m@ne\@citeo%
      \@tempcnta\@tempcntc\@tempcntb\@tempcntc\fi\fi}}\@citeo}{#1}}%

\def\@citeo{\ifnum\@tempcnta>\@tempcntb\else\@citea
  \def\@citea{,\penalty\@m}%
  \ifnum\@tempcnta=\@tempcntb\the\@tempcnta\else
   {\advance\@tempcnta\@ne\ifnum\@tempcnta=\@tempcntb \else
\def\@citea{--}\fi
    \advance\@tempcnta\m@ne\the\@tempcnta\@citea\the\@tempcntb}\fi\fi}

\catcode`\@=12

\global\mathchardef\lesssim "142E

\newcommand{\slL}{\raise.15ex\hbox{$/$}\kern-.53em\hbox{$L$}}
\newcommand{\slP}{\raise.15ex\hbox{$/$}\kern-.53em\hbox{$P$}}
\newcommand{\slp}{\raise.1ex\hbox{$/$}\kern-.63em\hbox{$p$}}
\newcommand{\slq}{\raise.1ex\hbox{$/$}\kern-.63em\hbox{$q$}}
\newcommand{\slv}{\raise.1ex\hbox{$/$}\kern-.63em\hbox{$v$}}
\newcommand{\slR}{\raise.15ex\hbox{$/$}\kern-.53em\hbox{$R$}}
\newcommand{\slQ}{\raise.15ex\hbox{$/$}\kern-.53em\hbox{$Q$}}
\newcommand{\slK}{\raise.15ex\hbox{$/$}\kern-.53em\hbox{$K$}}
\newcommand{\slk}{\raise.15ex\hbox{$/$}\kern-.53em\hbox{$k$}}
\newcommand{\slSigma}{\raise.15ex\hbox{$/$}\kern-.53em\hbox{$\Sigma$}}
\newcommand{\slcalP}{\raise.15ex\hbox{$/$}\kern-.63em\hbox{$\cal P$}}
\newcommand{\slA}{\raise.15ex\hbox{$/$}\kern-.73em\hbox{$A$}}
\newcommand{\slbfA}{\raise.15ex\hbox{$/$}\kern-.73em\hbox{${\imb A}$}}
\newcommand{\slpartial}{\raise.15ex\hbox{$/$}\kern-.53em\hbox{$\partial$}}

\newcommand{\be}{\begin{equation}}
\newcommand{\ee}{\end{equation}}
\newcommand{\bea}{\begin{eqnarray}}
\newcommand{\ena}{\end{eqnarray}}

\def\build#1\over#2{\mathrel{\mathop{\kern 0pt#1}\limits_{#2}}}

\font\tenimbf=cmmib10 at 10pt
\font\sevenimbf=cmmib10 at 7pt
\font\fiveimbf=cmmib10 at 5pt
\newfam\imbf
\textfont\imbf=\tenimbf
\scriptfont\imbf=\sevenimbf
\scriptscriptfont\imbf=\fiveimbf
\def\imb{\fam\imbf\tenimbf}

\begin{document}
\title{\bf{Probing colored glass\\ via $q\bar{q}$ photoproduction}\\
  II: diffraction}
\author{F.~Gelis$^{(1)}$, A.~Peshier$^{(2)}$}
\maketitle
\begin{center}
\begin{enumerate}
\item Laboratoire de Physique T\'eorique,\\
B\^at. 210, Universit\'e Paris XI,\\
91405 Orsay Cedex, France
\item Brookhaven National Laboratory,\\
Physics Department, Nuclear Theory,\\
Upton, NY-11973, USA
\end{enumerate}
\end{center}

\begin{abstract}
  In this paper, we consider the diffractive photoproduction of
  quark-antiquark pairs in peripheral heavy ion collisions. The color
  field of the nuclei is treated within the Colored Glass Condensate
  model. The cross-section turns out to be very sensitive to the
  value of the saturation scale.
\end{abstract}
\vskip 4mm
\centerline{\hfill BNL-NT-01/27, LPT-ORSAY-01-110}

\section{Introduction}
In recent years, a lot of work has been devoted to the understanding
of the parton distributions inside highly energetic hadrons or nuclei
(see \cite{Muell4,McLer1} for a pedagogical introduction). A central
question in this field is the issue of parton saturation at small
values of the momentum fraction $x$ \cite{GriboLR1,MuellQ1,FrankS1}.
Indeed, the BFKL equation \cite{Lipat1,KuraeLF1,BalitL1}, when applied
to extremely small values of $x$, leads to cross-sections that are too
large to be compatible with unitarity constrains. It has been argued
that this description becomes inadequate when non-linear effects in
QCD become important, which happens when the phase-space density
reaches the order of $1/\alpha_s$ \cite{GriboLR1,JalilKMW1,KovchM1,KovchM2}.

A theoretical description of saturation is provided by the model
introduced by McLerran and Venugopalan \cite{McLerV1,McLerV2,McLerV3},
in which small-$x$ gluons inside a fast moving nucleus are described by
a classical color field, which is motivated by the large occupation
number of the corresponding modes. The color field is determined by the
classical Yang-Mills equation with a current induced by the partons at
large values of $x$.  The distribution of hard color sources inside the
nucleus is determined by a functional density which is assumed to be
Gaussian in the original form of the model. This physical picture has
been named ``Colored Glass Condensate''. When quantum corrections are
taken into account, the density distribution evolves according to a
functional evolution equation
\cite{JalilKLW1,JalilKLW2,JalilKLW3,JalilKLW4,KovneM1,KovneMW3,Balit1,Kovch3,IancuLM1,IancuLM2}
which reduces in special cases to the BFKL equation. It is interesting
to note that this evolution equation was recently reobtained in the
dipole picture \cite{Muell5}, which is the natural description in the
frame where the nucleus is at rest.

The classical version of the color glass condensate model contains
only one parameter, called ``saturation scale'' $Q_s$, and defined as
the transverse momentum scale below which saturation effects start
being important. This scale increases with energy and with the size of
the nucleus \cite{Muell4,JalilKMW1,KovchM1}. Therefore, at energies
high enough so that $Q_s \gg \Lambda_{_{QCD}}$, the coupling constant
at the saturation scale is small, which makes perturbative approaches
feasible.

Note that another model inspired by saturation has also been developed
by Golec-Biernat and Wusthoff in order to fit the HERA data
\cite{GolecW1,GolecW2}. There, saturation is included via a
phenomenological formula for the dipole cross-section
$\sigma(r_\perp)=\sigma_0(1-\exp(-r_\perp^2/R(x)^2))$ that saturates for
large dipole sizes. The radius $R(x)$ is the dipole size above which the
saturation occurs, and is taken to be of the form
$R(x)=(x/x_0)^{\lambda/2}$GeV${}^{-1}$. A fit to inclusive data at small
$x$ gave $\sigma_0=23$mb, $x_0=3.10^{-4}$ and $\lambda=0.29$.  This
phenomenological model has also been successfully applied to the
description of diffractive HERA data \cite{GolecW3}. The main
differences between this model and the colored glass condensate model
are two-fold: in the model by Golec-Biernat and Wusthoff, saturation is
introduced by hand, and the $x$ evolution of the saturation scale is
also ad hoc. In the colored glass condensate, saturation emerges
explicitly as a consequence of solving the fully non-linear classical
Yang-Mills equations, and the quantum evolution is described by an
evolution equation that may lead in principle to an $x$ dependence much
more complicated than a power law.  One may also note the attempts
\cite{GotsmLLMT1,GotsmLLMT2} to fill the gap between the
phenomenological approach of Golec-Biernat and Wusthoff and perturbative
QCD, by using perturbative QCD as a starting point and adding shadowing
corrections {\sl a la} Glauber-Mueller.

Saturation effects, and in particular the color glass condensate model
can be tested in deep inelastic scattering experiments by measuring
structure functions like $F_2$ \cite{McLerV4}.  Interestingly, it also
leads to potentially observable effects in the field of heavy ion
collisions.  For central collisions, the colored glass condensate model
has been used to compute the distribution of the gluons produced
initially, both analytically \cite{KovneMW1,KovneMW2,Kovch4} and
numerically on a 2-dimensional lattice \cite{KrasnV1,KrasnV2}.  This
initial gluon multiplicity is expected to control directly the observed
final multiplicity. Recently, we have shown \cite{GelisP1} that the
parton saturation leads to interesting effects also in peripheral
collisions, where it can be probed via the photoproduction of
quark-antiquark pairs. In particular, the saturation scale $Q_s$
controls directly the position of the maximum in the transverse momentum
spectrum of the pairs.

\begin{figure}[ht]
\centerline{\resizebox*{!}{3cm}{\includegraphics{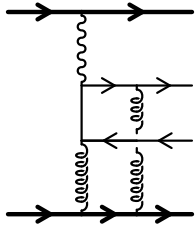}}}
\caption{\label{fig:proto}Prototype of the diagrams contributing to
the diffractive photoproduction of a $q\bar{q}$ pair in $AA$
collisions.}
\end{figure}

The present paper is an extension of this previous work, where we
study the {\sl diffractive} photoproduction of quark-antiquark pairs
in peripheral heavy ion collisions. By diffractive production, we mean
events in which the nuclei remain intact, and with a rapidity gap
between the nuclei and the produced objects characterizing the
topology of the final state. This implies that the exchange between
the color field and the quark pair is color singlet, and carries a
vanishing total transverse momentum. The simplest diagram for this
process is represented in figure \ref{fig:proto}.
We consider large ultra-relativistic nuclei, and we take into account
the electromagnetic interaction to lowest order in the coupling (or,
equivalently, in the leading logarithmic approximation), and the
interactions with the strong color background field to all orders. We
find that the cross-section of this process is very sensitive to the
value of the saturation scale in the colored glass condensate. Note that
we use a Gaussian distribution in order to describe the hard color
sources that are generating the classical color fields. This means that
no quantum evolution has been included in this calculation.  It is known
that in the colored glass condensate model, the quantum corrections lead
to a modification of the distribution function of hard sources when the
energy increases
\cite{JalilKLW1,JalilKLW2,JalilKLW3,JalilKLW4,KovneM1,KovneMW3,Balit1,Kovch3,IancuLM1,IancuLM2}
(a similar evolution equation has been derived by Kovchegov and Levin in
the dipole language \cite{KovchL1}, and has been recently shown to be
equivalent to the evolution equation in the colored glass condensate by
Mueller \cite{Muell5}). In general, this distribution will not keep a
Gaussian form due to these corrections.  However, in order to assess
very simply saturation effects in diffractive ultra-peripheral
collisions, we do not include this refinement here. Note that in
principle, extending our calculation to include quantum evolution is
just a matter of changing the correlator $\left<U(0)U^\dagger({\imb
x}_\perp)\right>_{\overline{\rho}}$ that appears in our results (see
Eq.~(\ref{eq:def-C}) for instance) and contains all the information
about the properties of the colored glass condensate by one that has
been calculated with the evolved distribution of ``hard sources''.

The structure of this paper is as follows. In section \ref{sec:P1}, we
summarize results relevant for the calculation of particle production
in a classical background field. We also recall how to
calculate diffractive quantities in the McLerran-Venugopalan model. In
section \ref{sec:lowest-order}, we compute the pair production
amplitude to lowest order in the electromagnetic coupling constant.
In section \ref{sec:kt-spectrum}, an expression for the transverse
momentum spectrum of the components of the pair is derived, and in
section \ref{sec:cross-section} results for the integrated diffractive
cross-section are presented.  Finally, section \ref{sec:conclusion} is
devoted to concluding remarks.

\section{Reduction formulas for pair production}
\label{sec:P1}
In this section, we remind the reader of reduction formulas one can
use to compute various physical quantities in a classical background
field. The formalism was derived in \cite{BaltzGMP1}, and here we are
only going to recall the results, using similar notations. Different
from the situation studied in \cite{BaltzGMP1}, the relevant classical
background field is now a superposition of an electromagnetic field
and a color field. The color field is associated to the covariant
gauge distribution of hard sources $\widetilde{\rho}$ (see section 2
of \cite{GelisP1}). In order to calculate physical quantities, one has
to average over an ensemble distribution of these sources. Quantities
obtained before averaging over the sources will be denoted by an index
$\widetilde{\rho}$.

\subsection{$\overline{n}$ versus $P_1$}
Certain quantities of interest for the problem of quark-antiquark
production in a background field can be related to quark propagators
with different boundary conditions in this background field
\cite{BaltzGMP1}. The main results are the following.

The probability to produce {\sl exactly one} $q\bar{q}$ pair in a
background field can be related to the {\sl Feynman} (or time-ordered)
propagator of a quark via
\begin{equation}
P_1[\widetilde{\rho}]=|\left<0_{\rm out}|0_{\rm in}\right>_{[\widetilde{\rho}]}|^2\,
\int
{{d^3{\imb q}}\over{(2\pi)^3 2\omega_{\imb q}}}
{{d^3{\imb p}}\over{(2\pi)^3 2\omega_{\imb p}}}
\left|
\overline{u}({\imb q})
{\cal T}_F^{[\widetilde{\rho}]}(q,-p)
v({\imb p})
\right|^2\; .
\label{eq:P1-def}
\end{equation}
In this equation, ${\cal T}_F^{[\widetilde{\rho}]}$ is the interacting
part of the Feynman propagator of a quark in the background field
generated by the source $\widetilde{\rho}$. In terms of the exact
Feynman propagator $G_F^{[\widetilde{\rho}]}$ and the free Feynman
propagator $G_F^0$, it is defined as
\begin{equation}
G_F^{[\widetilde{\rho}]}=G_F^0+G_F^0 {\cal T}_F^{[\widetilde{\rho}]} G_F^0\; .
\end{equation}
Note in Eq.~(\ref{eq:P1-def}) the presence of the square of the
vacuum-to-vacuum amplitude
$\left<0_{\rm out}|0_{\rm in}\right>_{[\widetilde{\rho}]}$, which is
essential for the unitarity of the result\footnote{In a field theory
  with a time-dependent background field, the vacuum-to-vacuum
  amplitude is not a pure phase, but has a modulus different from one.
  This is a mere consequence of the fact that such a background field
  can produce or destroy particles.}.

Conversely, the {\sl average number} of $q\bar{q}$ pairs produced in a
background field, $\bar{n} = \sum n P_n$, where $P_n$ are the
production probabilities, can be expressed in terms of the interacting
part of the {\sl retarded} propagator of a quark in this background:
\begin{equation}
\overline{n}[\widetilde{\rho}]=
\int
{{d^3{\imb q}}\over{(2\pi)^3 2\omega_{\imb q}}}
{{d^3{\imb p}}\over{(2\pi)^3 2\omega_{\imb p}}}
\left|
\overline{u}({\imb q})
{\cal T}_R^{[\widetilde{\rho}]}(q,-p)
v({\imb p})
\right|^2\; .
\label{eq:nbar-def}
\end{equation}
Another difference to Eq.~(\ref{eq:P1-def}) is the absence of the
vacuum-to-vacuum factor.

For the background field of two nuclei in peripheral collisions, an
interesting aspect is that in the ultra-relativistic limit the
retarded propagator has a much simpler structure than the Feynman
propagator \cite{BaltzGMP1}. This enables one to write the retarded
propagator of a quark explicitly, while this is not possible for the
time-ordered one.

\subsection{Diffractive and inclusive quantities}
So far, we have given the expression for quantities that would be
observed for the background field configuration associated to a
specific source $\widetilde{\rho}$. However, this is not what is
measured in experiments where quantities are usually averaged over
many collisions. In the colored glass condensate model, this
translates into an average over the sources, which are assumed to have
a Gaussian distribution in the classical version of the model.

The prescription to average observables, like for instance $P_1$ or
$\overline{n}$, over $\widetilde{\rho}$ is
\begin{equation}
\left<{\cal O}\right>_{\widetilde{\rho}}\equiv \int [d\widetilde{\rho}]w[\widetilde{\rho}]{\cal O}[\widetilde{\rho}]\; ,
\end{equation}
with the distribution
\begin{equation}
w[\widetilde{\rho}]\equiv \exp\left\{
-\int dx^-d^2{\imb x}_\perp {{\widetilde{\rho}_a(x^-,{\imb x}_\perp)\widetilde{\rho}^a(x^-,{\imb x}_\perp)}\over{2\mu^2(x^-)}}
\right\}\; .
\end{equation}
This averaging procedure yields inclusive physical quantities,
``inclusive'' meaning here that the final state of the nucleus which
interacts by its color field is not observed. For the present case of
$q\bar{q}$ photoproduction, it means that the process under
consideration is: $ZA\to Z q\bar{q} X$ (the nucleus interacting
electromagnetically, denoted by $Z$, remains intact, while the other
nucleus may fragment).

\begin{figure}[ht]
\centerline{\resizebox*{!}{3.5cm}{\includegraphics{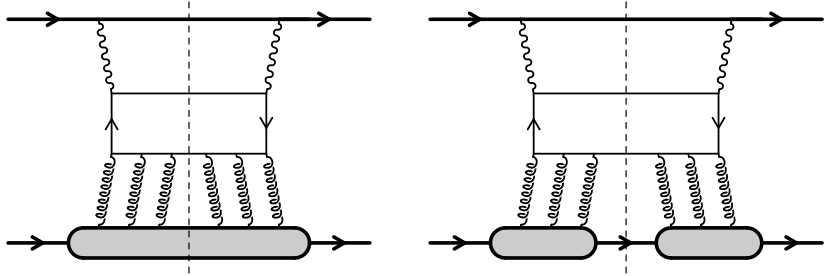}}}
\caption{\label{fig:diffrac}Inclusive (left) vs. diffractive (right)
photoproduction of $q\bar{q}$ pairs. The shaded blob represents the
average over a hard color source. For the diffractive cross-section,
one has to average the amplitude before squaring it.}
\end{figure}

It was also noted that one can obtain a diffractive quantity (i.e.\ for
events where the nucleus $A$ does not break up) by averaging the {\sl
amplitude} over the sources $\widetilde{\rho}$ before squaring it in
order to get a probability. This procedure has been outlined in
\cite{BuchmGH1,BuchmMH1} and justified in \cite{KovchM1}\footnote{For
another discussion of the averaging procedure we refer the reader to
\cite{KovneW1}.}. The difference between the two ways of performing the
average is illustrated in figure \ref{fig:diffrac}, where the shaded
blob indicates an average over the color sources.
This average has the property that the gluons attached to one such
blob carry a zero total transverse momentum and no net color. Given
this, it is intuitive that the nucleus remains intact in the final
state (i.e.\ on the cut represented by a dashed line) for the right
diagram. In addition, this condition leads to a rapidity gap between
the nucleus and the $q\bar{q}$ pair.  More formal considerations on
this issue can be found in \cite{KovneW1}.

Applied to the production of one $q\bar{q}$ pair, the inclusive
probability is given by
\begin{equation}
P_1^{\rm incl}=
\int
{{d^3{\imb q}}\over{(2\pi)^3 2\omega_{\imb q}}}
{{d^3{\imb p}}\over{(2\pi)^3 2\omega_{\imb p}}}
\left<|\left<0_{\rm out}|0_{\rm in}\right>_{[\widetilde{\rho}]}|^2\,\left|\,
\overline{u}({\imb q})
{\cal T}_F^{[\widetilde{\rho}]}(q,-p)
v({\imb p})
\right|^2\right>_{\widetilde{\rho}}\; .
\label{eq:P1-incl-def}
\end{equation}
Note that the vacuum-to-vacuum factor must be included inside the
average, because it depends on the source $\widetilde{\rho}$. On the
other hand, the diffractive probability is given by
\begin{equation}
P_1^{\rm diff}=
\int
{{d^3{\imb q}}\over{(2\pi)^3 2\omega_{\imb q}}}
{{d^3{\imb p}}\over{(2\pi)^3 2\omega_{\imb p}}}
\left|
\left<
\left<0_{\rm out}|0_{\rm in}\right>_{[\widetilde{\rho}]}
\overline{u}({\imb q})
{\cal T}_F^{[\widetilde{\rho}]}(q,-p)
v({\imb p})
\right>_{\widetilde{\rho}}
\right|^2\; .
\label{eq:P1-diff-def}
\end{equation}
Regarding, on the other hand, the average number of pairs in
diffractive events, it is important to realize that this quantity
cannot be obtained from the expression (\ref{eq:nbar-def}), which
is already a combination of probabilities (i.e.\ squared amplitudes).
From Eq.~(\ref{eq:nbar-def}), therefore, only the average number of
pairs produced in a collision, regardless of what happens to the
nucleus, can be calculated \cite{GelisP1}. In the rest of this paper,
we focus on the calculation of $P_1^{\rm diff}$.

\section{Expression of $P_1$ to order $(Z\alpha)^2$}
\label{sec:lowest-order}
The results of the previous section are valid to all orders in the
coupling constants occurring in the problem, in particular the
electromagnetic coupling $Z\alpha e_q$ of a quark with charge $e_q$
(in units of the positron charge) to the nucleus with atomic number
$Z$. However, the Feynman propagator and the vacuum-to-vacuum
amplitude are not known to all orders in $Z\alpha$, even in the
ultra-relativistic limit (see \cite{BaltzGMP1}). In this section, we
show that results can be obtained in closed form if one works at the
lowest non-trivial order in $Z\alpha$, but to all orders in the
interactions with the strong color field of the other nucleus.
Although the value of $Z\alpha$ might seem too large ($Z\alpha\approx
0.6$ for a gold nucleus) as a useful expansion parameter, it turns out
that at leading order the production probabilities contain large
logarithms whose argument is the center of mass energy of the
collision. It can be shown that higher order terms in $Z\alpha$ are
suppressed by at least one power of this large logarithm, which
justifies the approximation for large enough collision energies.

\subsection{Pair production amplitude}
\begin{figure}[ht]
\centerline{\resizebox*{!}{2cm}{\includegraphics{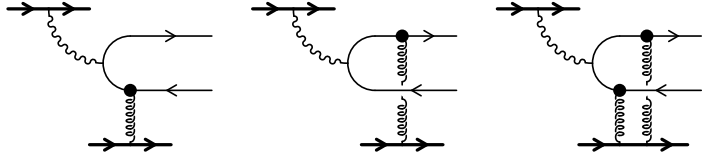}}}
\caption{\label{fig:diagrams-M1}The three diagrams contributing to ${\cal
    T}_F$ at lowest order in the electromagnetic coupling constant.
  The black dot denotes the time-ordered eikonal matrix $T_F$ that describes
  the interaction of a quark or antiquark with the colored glass
  condensate.}
\end{figure}

Many simplifications occur when one considers only the leading order in
$Z\alpha$. Indeed, since
\begin{eqnarray}
&&\left|\overline{u}({\imb q}) {\cal
T}_F^{[\widetilde{\rho}]}(q,-p)v({\imb p}) \right|^2 = {\cal
O}((Z\alpha)^2) \; , \nonumber\\
&&|\left<0_{\rm out}|0_{\rm
in}\right>_{\widetilde{\rho}}|^2=1+{\cal O}((Z\alpha)^2)\; ,
\end{eqnarray}
one can approximate the vacuum-to-vacuum factor by $1$ in the
expression for $P_1$ at leading order\footnote{A word of caution is
  required here. The vacuum-to-vacuum factor is necessary for
  unitarity, and this approximation can lead to probabilities that are
  larger than one. This happens because the ``smallness'' of
  $(Z\alpha)^2$ can be overcompensated by the large logarithms that
  accompany it. However, this will not be a problem as long as the
  multiplicity $\overline{n}$ remains much smaller than unity, as
  discussed later.}  in $Z\alpha$. Another notable simplification
occurs in the Feynman propagator of the quark in the background field.
From an infinite series, it simplifies to the sum of only three terms
that are represented in figure \ref{fig:diagrams-M1}. The interaction
with the color field of the second nucleus is accounted for to all
orders by eikonal factors depicted by the black dot.
Thus, one can explicitly write down the time-ordered scattering matrix
${\cal T}_F^{[\widetilde{\rho}]}$ in terms of the individual
scattering matrices off the two nuclei. The sum of the three diagrams
of figure \ref{fig:diagrams-M1} is formally
\begin{equation}
{\cal T}_F^{[\widetilde{\rho}]}=
T_F^{^{QCD}}G_F^0 T_F^{^{QED}}
+ T_F^{^{QED}}G_F^0 T_F^{^{QCD}}
+ T_F^{^{QCD}}G_F^0 T_F^{^{QED}} G_F^0 T_F^{^{QCD}}\; ,
\label{eq:TF}
\end{equation}
where $G_F^0$ is the free time-ordered quark propagator:
\begin{equation}
G_F^0(p)=i{{\slp+m}\over{p^2-m^2+i\epsilon}}\; .
\end{equation}
$T_F^{^{QCD}}$ is the time-ordered scattering matrix of a quark
in the color field of the second nucleus:
\begin{equation}
T_F^{^{QCD}}(q,p)=2\pi\delta(p^--q^-)\slv_+ {\rm sign}(p^-)
\int d^2{\imb x}_\perp \left[U({\imb x}_\perp)^{{\rm sign}(p^-)}-1\right]
e^{i({\imb q}_\perp-{\imb p}_\perp)\cdot{\imb x}_\perp}
\end{equation}
with
\begin{equation}
U({\imb x}_\perp)\equiv {\rm T}\exp\left\{ -ig^2 \int_{-\infty}^{+\infty}dz^-{1\over{\nabla_\perp^2}}\widetilde{\rho}_a(
z^-,{\imb x}_\perp)t^a\right\}\; ,
\end{equation}
where the $t^a$'s are the generators of the fundamental representation
of $SU(N)$. At lowest order in $Z\alpha$, the scattering matrix of the
quark in the electromagnetic field of the first nucleus simplifies to
\begin{equation}
T_F^{^{QED}}(q,p)=2\pi\delta(p^+-q^+)\slv_-
{{4\pi Z\alpha e_q}\over{({\imb q}_\perp-{\imb p}_\perp)^2}} e^{i({\imb q}_\perp-{\imb p}_\perp)\cdot{\imb b}}\; ,
\end{equation}
where ${\imb b}$ is the impact parameter in the collision of the
two nuclei (we chose the origin of coordinates in the transverse plane
so that all the ${\imb b}$ dependence goes into the electromagnetic
scattering matrix).

From there, it is easy to work out a compact expression for the
amplitude to create a $q\bar{q}$ pair by photoproduction in a
peripheral collision. Replacing by $1$ the factor $\left<0_{\rm
in}|0_{\rm out}\right>_{[\widetilde{\rho}]}$ at this order in
$Z\alpha$, this amplitude is
\begin{equation}
M_1^{[\widetilde{\rho}]}({\imb p},{\imb q})=
\overline{u}({\imb q})
{\cal T}_F^{[\widetilde{\rho}]}(q,-p)v({\imb p})\; .
\end{equation}
For the first two terms in Eq.~(\ref{eq:TF}), all the integrals over
the $\pm$ components of the momentum transfer can be done thanks to
the delta functions present in the individual scattering matrices. For
the third term, there are two independent transfer momenta. Three of
the $\pm$ integrals can still be performed with the delta functions,
and the fourth one can be calculated in the complex plane with the
theorem of residues. We obtain:
\begin{eqnarray}
&&\!\!\!\!\!\!M_1^{[\widetilde{\rho}]}({\imb p},{\imb q})=\nonumber\\
&&i\int {{d^2{\imb k}_\perp}\over{(2\pi)^2}} F^{^{QED}}({\imb p}_\perp+{\imb q}_\perp-{\imb k}_\perp) F^{^{QCD}}_+({\imb k}_\perp)
{{\overline{u}({\imb q})\slv_+(\slq-\slk+m)\slv_- v({\imb p})}\over
{2p^+q^-+({\imb q}_\perp-{\imb k}_\perp)^2+m^2}}\nonumber\\
&&\!\!\!\!+i\int {{d^2{\imb k}_\perp}\over{(2\pi)^2}} F^{^{QED}}({\imb p}_\perp+{\imb q}_\perp-{\imb k}_\perp) F^{^{QCD}}_-({\imb k}_\perp)
{{\overline{u}({\imb q})\slv_-(\slk-\slp+m)\slv_+ v({\imb p})}\over
{2p^-q^++({\imb p}_\perp-{\imb k}_\perp)^2+m^2}}\nonumber\\
&&\!\!\!\!+i\int {{d^2{\imb k_1}_\perp}\over{(2\pi)^2}}
{{d^2{\imb k_2}_\perp}\over{(2\pi)^2}}
F^{^{QED}}({\imb p}_\perp+{\imb q}_\perp-{\imb k_1}_\perp-{\imb k_2}_\perp)
F^{^{QCD}}_+({\imb k_1}_\perp)F^{^{QCD}}_-({\imb k_2}_\perp)\nonumber\\
&&\qquad\qquad\times{{{\overline{u}({\imb q})
\slv_+(\slq-\slk_1+m)\slv_-(\slk_2-\slp+m)\slv_+
v({\imb p})}
\over{2q^-(({\imb p}_\perp-{\imb k_2}_\perp)^2+m^2)+2p^-(({\imb q}_\perp-{\imb k_1}_\perp)^2+m^2)}}}\; ,
\end{eqnarray}
where we denote
\begin{eqnarray}
&&F^{^{QED}}({\imb k}_\perp)
\equiv {{4\pi Z\alpha}\over{{\imb k}_\perp^2}}
e^{i{\imb k}_\perp\cdot{\imb b}}
\; ,\nonumber\\
&&F^{^{QCD}}_\pm({\imb k}_\perp)\equiv\int d^2{\imb x}_\perp
e^{i{\imb k}_\perp\cdot{\imb x}_\perp} \Big(U^{\pm 1}({\imb x}_\perp)-1\Big)\; .
\end{eqnarray}
At this stage, it is a simple matter of algebra to combine the three
terms into a single one:
\begin{eqnarray}
&&\!\!\!\!\!\!M_1^{[\widetilde{\rho}]}({\imb p},{\imb q})=\nonumber\\
&&i\int {{d^2{\imb k_1}_\perp}\over{(2\pi)^2}}
{{d^2{\imb k_2}_\perp}\over{(2\pi)^2}}
F^{^{QED}}({\imb p}_\perp+{\imb q}_\perp-{\imb k_1}_\perp-{\imb k_2}_\perp)
\nonumber\\
&&\!\!\!\!\!\!\times
\int d^2{\imb x_1}_\perp d^2{\imb x_2}_\perp
e^{i{\imb k_1}_\perp\cdot{\imb x_1}_\perp}
e^{i{\imb k_2}_\perp\cdot{\imb x_2}_\perp}
\Big[U({\imb x_1}_\perp)U^\dagger({\imb x_2}_\perp)-1\Big]M({\imb p},{\imb q}|{\imb k_1}_\perp,{\imb k_2}_\perp)\; ,\nonumber\\
&&
\label{eq: Mto}
\end{eqnarray}
where we define:
\begin{eqnarray}
&&M({\imb p},{\imb q}|{\imb k_1}_\perp,{\imb
k_2}_\perp)\equiv\nonumber\\
&&\qquad\equiv
\overline{u}({\imb q}){{{
\slv_+(\slq-\slk_1+m)\slv_-(\slk_2-\slp+m)\slv_+
}
\over{2q^-(({\imb p}_\perp-{\imb k_2}_\perp)^2+m^2)+2p^-(({\imb q}_\perp-{\imb k_1}_\perp)^2+m^2)}}}v({\imb p})\; .
\end{eqnarray}
This is the formula we will use in the following in order to compute
the diffractive cross-section for the photoproduction of a
quark-antiquark pair.

\subsection{Equality of $P_1^{\rm incl}$ and $\overline{n}$ at order $(Z\alpha)^2$}
Before going to the detailed calculation of the diffractive
cross-section, it is interesting to check a property that holds at
order $(Z\alpha)^2$. At this order, we have:
\begin{eqnarray}
&&\overline{n}={\cal O}((Z\alpha)^2)\nonumber\\
&&P_1^{\rm incl}={\cal O}((Z\alpha)^2)\nonumber\\
&&P_n^{\rm incl}=0\qquad{\rm for\ }n\ge 2\; ,
\end{eqnarray}
where $P_n^{\rm incl}$ is the inclusive probability to produce $n$
pairs in a collision. Since by definition $\overline{n}=P_1^{\rm
  incl}+2 P_2^{\rm incl}+\cdots$, we immediately conclude that at
order $(Z\alpha)^2$, one has $\overline{n}=P_1^{\rm incl}$. Although
this property is obvious at such a formal level, it is interesting to
see how it comes about in the explicit calculation.  Indeed, we know
that $\overline{n}$ is given by the square of a retarded amplitude, to
which only the first two diagrams of figure \ref{fig:diagrams-M1}
contribute \cite{GelisP1}, while all three diagrams contribute to
$P_1^{\rm incl}$.  The contribution of the first two diagrams to the
retarded amplitude was calculated in \cite{GelisP1},
\begin{eqnarray}
&&\!\!\!\!\!\!M_R^{[\widetilde{\rho}]}({\imb p},{\imb q})=\nonumber\\
&&-i\int{{d^2{\imb k}_\perp}\over{(2\pi)^2}}
F^{^{QED}}({\imb p}_\perp+{\imb q}_\perp-{\imb k}_\perp)\int d^2{\imb x}_\perp e^{i{\imb k}_\perp\cdot{\imb x}_\perp}
\Big[U({\imb x}_\perp)-1\Big]\nonumber\\
&&\times\overline{u}({\imb q})\Big\{
{{\slv_+(\slq-\slk+m)\slv_- v({\imb p})}\over
{2p^+q^-+({\imb q}_\perp-{\imb k}_\perp)^2+m^2}}
+{{\slv_-(\slk-\slp+m)\slv_+ v({\imb p})}\over
{2p^-q^++({\imb p}_\perp-{\imb k}_\perp)^2+m^2}}
\Big\}v({\imb p})\; .\nonumber\\
&&
\end{eqnarray}
After some algebra, this expression can be rewritten as
\begin{eqnarray}
&&\!\!\!\!\!\!M_R^{[\widetilde{\rho}]}({\imb p},{\imb q})=\nonumber\\
&&i\int {{d^2{\imb k_1}_\perp}\over{(2\pi)^2}}
{{d^2{\imb k_2}_\perp}\over{(2\pi)^2}}
F^{^{QED}}({\imb p}_\perp+{\imb q}_\perp-{\imb k_1}_\perp-{\imb k_2}_\perp)
\nonumber\\
&&\times
\int d^2{\imb x_1}_\perp d^2{\imb x_2}_\perp
e^{i{\imb k_1}_\perp\cdot{\imb x_1}_\perp}
e^{i{\imb k_2}_\perp\cdot{\imb x_2}_\perp}
\Big[U({\imb x_1}_\perp)-U({\imb x_2}_\perp)\Big]M({\imb p},{\imb q}|{\imb k_1}_\perp,{\imb k_2}_\perp)\; .\nonumber\\
&&
\end{eqnarray}
Therefore, we see that at this order in $Z\alpha$, the time-ordered
amplitude (\ref{eq: Mto}) and the retarded amplitude differ only by a
factor $U({\imb x_2}_\perp)$ in the integrand. When squaring the
amplitudes in order to get either $\overline{n}$ or $P_1^{\rm incl}$,
this results in a factor $U({\imb x_2}_\perp)U^\dagger({\imb
  x_2}^\prime_\perp)$. Integrating over the transverse momentum ${\imb
  p}_\perp$ of the antiquark yields $\delta({\imb x_2}_\perp-{\imb
  x_2}^\prime_\perp)$, and this factor becomes unity ($U$ is a unitary
matrix), leading to the expected equality $\overline{n}=P_1^{\rm
  incl}$.

\subsection{A useful inequality}
There is another very important property that we can prove without
performing explicitly all the integrals. Performing only the integrals
over ${\imb p}_\perp$ and ${\imb q}_\perp$, one can see that the eikonal
color matrices present in the square of the time-ordered amplitude are
of the form
\begin{eqnarray}
&&\left<\Big[U({\imb x_1}_\perp)U^\dagger({\imb x_2}_\perp)-1\Big]
\Big[U({\imb x_2}_\perp)U^\dagger({\imb x_1}_\perp)-1\Big]\right>_{\widetilde{\rho}}\nonumber\\
&&\qquad\qquad=2\Big[1-\left<U({\imb x_1}_\perp)U^\dagger({\imb x_2}_\perp)\right>_{\widetilde{\rho}}\Big]
\end{eqnarray}
in the inclusive probability, and
\begin{eqnarray}
&&\left<\Big[U({\imb x_1}_\perp)U^\dagger({\imb x_2}_\perp)-1\Big]\right>_{\widetilde{\rho}}
\left<\Big[U({\imb x_2}_\perp)U^\dagger({\imb x_1}_\perp)-1\Big]\right>_{\widetilde{\rho}}\nonumber\\
&&\qquad\qquad=\Big[1-\left<U({\imb x_1}_\perp)U^\dagger({\imb x_2}_\perp)\right>_{\widetilde{\rho}}\Big]^2
\end{eqnarray}
in the diffractive probability. In the above equations, we have used
the fact that the correlator $\left<U({\imb x_1}_\perp)U^\dagger({\imb
x_2}_\perp)\right>_{\widetilde{\rho}}$ is a real number, and symmetric
under the exchange of the labels $1$ and $2$. All the other factors in
the integrand are the same in both probabilities.

Recalling now the fact that $0\le\left<U({\imb x_1}_\perp)U^\dagger({\imb
    x_2}_\perp)\right>_{\widetilde{\rho}}\le 1$ (see the appendix of
\cite{GelisP1}), we also know that
\begin{equation}
\Big[1-\left<U({\imb x_1}_\perp)U^\dagger({\imb x_2}_\perp)\right>_{\widetilde{\rho}}\Big]^2\le \Big[1-\left<U({\imb x_1}_\perp)U^\dagger({\imb x_2}_\perp)\right>_{\widetilde{\rho}}\Big]\; .
\end{equation}
From this it is easy to infer
the inequality
\begin{equation}
P_1^{\rm diff} \le \frac12\, P_1^{\rm incl}\, ,
\label{eq:inequal}
\end{equation}
which will provide a useful consistency check at the end of the
calculation.
Using the fact that $\left< U({\imb x_1}_\perp) U^\dagger({\imb
x_2}_\perp) \right>_{\widetilde{\rho}}$ vanishes in the limit
$Q_s\to +\infty$, we also obtain from the above considerations
\begin{equation}
\lim_{Q_s\to+\infty} P_1^{\rm diff}={{P_1^{\rm incl}}\over 2}\; .
\end{equation}
Therefore, the upper bound for $P_1^{\rm diff}$ (the `black-disc
limit') is reached for infinite $Q_s$, i.e.\ in the asymptotic limit
where the center of mass energy of the collision is infinite.

\section{$p_\perp$ spectrum for diffraction}
\label{sec:kt-spectrum}
We are now in a position to derive an expression for the quantity
$P_1^{\rm diff}$. For this diffractive probability, we need to
evaluate the average over the sources $\widetilde{\rho}$ of the
amplitude $M_1^{[\widetilde{\rho}]}({\imb p},{\imb q})$.  This
involves the average $\left<U({\imb x_1}_\perp)U^\dagger({\imb
    x_2}_\perp)-1\right>_{\widetilde{\rho}}$, which was calculated in
the appendix of \cite{GelisP1},
\begin{eqnarray}
&&\left<U({\imb x_1}_\perp)U^\dagger({\imb
x_2}_\perp)-1\right>_{\widetilde{\rho}}\nonumber\\
&&\qquad={\cal P}({\imb x_1}_\perp){\cal P}({\imb x_2}_\perp)
\Big(e^{-B_2({\imb x_1}_\perp-{\imb x_2}_\perp)}-1\Big)
\nonumber\\
&&\qquad+({\cal P}({\imb x_1}_\perp)+{\cal P}({\imb x_2}_\perp)-2{\cal P}({\imb x_1}_\perp){\cal P}({\imb x_2}_\perp))\Big(e^{-B_1}-1\Big)\; .
\label{eq:average}
\end{eqnarray}
Here the function ${\cal P}({\imb x}_\perp)$ describes the transverse
profile of the nucleus (see \cite{GelisP1} for more details on how it
is introduced in the formalism). We assume ${\cal P}({\imb x}_\perp) =
0$ outside the nucleus, and ${\cal P}({\imb x}_\perp) = 1$ inside, and
that for large nuclei the transition at the surface of the nucleus
occurs over a distance much smaller than the nuclear radius $R$. In
the above equation, we denote:
\begin{eqnarray}
B_2({\imb x}_\perp)&\equiv&
Q_s^2\int d^2{\imb z}_\perp [G_0({\imb z}_\perp)-G_0({\imb z}_\perp-{\imb x}_\perp)]^2
\nonumber\\
&\approx& {{Q_s^2 x_\perp^2}\over{4\pi}}\ln\left({1\over{x_\perp \Lambda_{_{QCD}}}}\right) , \nonumber\\
B_1&\equiv& Q_s^2 \int d^2{\imb z}_\perp G_0({\imb z}_\perp)^2 \sim {{Q_s^2}\over{\Lambda_{_{QCD}}^2}}\gg 1
\end{eqnarray}
with $G_0({\imb x}_\perp)$ the 2-dimensional free
propagator\footnote{This propagator satisfies: ${\partial^2}G_0({\imb
x}_\perp-{\imb z}_\perp)/{\partial{\imb z}^2_\perp}=\delta({\imb
x}_\perp-{\imb z}_\perp)$.} and where $Q_s$ is the sa\-tu\-ra\-tion
mo\-men\-tum. In the previous equations, one has to introduce by hand
the scale $\Lambda_{_{QCD}}$ as a regulator for infrared
singularities. More details on this can be found in \cite{LamM1}.

In Eq.~(\ref{eq:average}), the second term depends on ${\imb
x_1}_\perp$ and ${\imb x_2}_\perp$ only via the profile function ${\cal
P}$. Performing explicitly the integral over these coordinates,
the contribution of this term to $\left<
M_1^{[\widetilde{\rho}]}({\imb p},{\imb q})
\right>_{\widetilde{\rho}}$ is:
\begin{eqnarray}
&&i\int {{d^2{\imb k_1}_\perp}\over{(2\pi)^2}}
{{d^2{\imb k_2}_\perp}\over{(2\pi)^2}}
F^{^{QED}}({\imb p}_\perp+{\imb q}_\perp-{\imb k_1}_\perp-{\imb k_2}_\perp)
M({\imb p},{\imb q}|{\imb k_1}_\perp,{\imb k_2}_\perp)\nonumber\\
&&\qquad\times
\left(\widetilde{\cal P}({\imb k_1}_\perp)(2\pi)^2\delta({\imb k_2}_\perp)
+\widetilde{\cal P}({\imb k_2}_\perp)(2\pi)^2\delta({\imb k_1}_\perp)
-2\widetilde{\cal P}({\imb k_1}_\perp)\widetilde{\cal P}({\imb k_2}_\perp)
\right)\; ,\nonumber\\
&&
\end{eqnarray}
where the function $\widetilde{\cal P}({\imb k}_\perp)$ is the Fourier
transform of the profile ${\cal P}({\imb x}_\perp)$. This function is
strongly peaked around ${\imb k}_\perp=0$, with a typical width of
order $1/R$. Therefore, both ${\imb k_1}_\perp$ and ${\imb k_2}_\perp$
are at most of the order of $1/R$ in this integral.  Since the factor
$M({\imb p},{\imb q}|{\imb k_1}_\perp,{\imb k_2}_\perp)$ does not
contain any scale as small as $1/R$ (this amplitude is controlled by
the scale of the quark mass $m$), it is legitimate to replace it by
the constant $M({\imb p},{\imb q}|0,0)$.  By choosing ${\imb
  k}_\perp\equiv {\imb k_1}_\perp+{\imb k_2}_\perp$ as one of the
integration variables, this contribution can be rewritten as
\begin{eqnarray}
&&2iM({\imb p},{\imb q}|0,0)
\int {{d^2{\imb k}_\perp}\over{(2\pi)^2}}
F^{^{QED}}({\imb p}_\perp+{\imb q}_\perp-{\imb k}_\perp)
\nonumber\\
&&\qquad\times\int{{d^2{\imb k}^\prime_\perp}\over{(2\pi)^2}}
\left(\widetilde{\cal P}({\imb k}_\perp)(2\pi)^2\delta({\imb k}^\prime_\perp)
-\widetilde{\cal P}({\imb k}_\perp)\widetilde{\cal P}({\imb k}_\perp-{\imb k}^\prime_\perp)
\right)\; .\nonumber\\
&&
\end{eqnarray}
At this point, it is trivial to see that the function of ${\imb
  k}_\perp$ appearing on the second line of the previous expression is
vanishing if we have ${\cal P}({\imb x}_\perp){\cal P}({\imb
  x}_\perp)\approx {\cal P}({\imb x}_\perp)$. This is indeed the case
if the transition of ${\cal P}({\imb x}_\perp)$ from 0 to 1 is sharp,
i.e.\ if we can neglect surface effects in the description of the
nucleus. To recapitulate, one can drop the second term of
Eq.~(\ref{eq:average}) if we have $m\gg 1/R$ and $R\gg
\Lambda_{_{QCD}}^{-1}$.

For the remaining term, it is convenient to resort to an approximation
already used in \cite{GelisP1}, valid when $Q_s\gg \Lambda_{_{QCD}}
\gg 1/R$:
\begin{equation}
{\cal P}({\imb x_1}_\perp){\cal P}({\imb x_2}_\perp)
e^{-B_2({\imb x_1}_\perp-{\imb x_2}_\perp)} \approx
{\cal P}({\imb x_1}_\perp)
e^{-B_2({\imb x_1}_\perp-{\imb x_2}_\perp)}\; .
\end{equation}
This approximation just indicates that due to the factor $\exp(-B_2)$
the difference ${\imb x_1}_\perp-{\imb x_2}_\perp$ is much smaller
than the nuclear radius. Additionally, using the same arguments as
above, the term ${\cal P}({\imb x_1}_\perp){\cal P}({\imb x_2}_\perp)$
can be replaced by a single factor ${\cal P}({\imb x_1}_\perp)$.
The Fourier transform of the first term of Eq.~(\ref{eq:average}) is
then trivial, as we have
\begin{eqnarray}
&&\int d^2{\imb x_1}_\perp d^2{\imb x_2}_\perp e^{i{\imb
k_1}_\perp\cdot{\imb x_1}_\perp} e^{i{\imb k_2}_\perp\cdot{\imb
x_2}_\perp} {\cal P}({\imb x_1}_\perp) \left(e^{-B_2({\imb
x_1}_\perp-{\imb x_2}_\perp)}-1\right)\nonumber\\
&&\qquad\qquad =\widetilde{\cal P}({\imb
k_1}_\perp+{\imb k_2}_\perp) (C({\imb k_2}_\perp)-(2\pi)^2\delta({\imb
k_2}_\perp))\; ,
\end{eqnarray}
where, like in \cite{GelisP1}, we denote
\begin{equation}
C({\imb k}_\perp)\equiv \int d^2{\imb x}_\perp
e^{i{\imb k}_\perp\cdot{\imb x}_\perp}
\left<U(0)U^\dagger({\imb x}_\perp)\right>_{\widetilde{\rho}}\; .
\label{eq:def-C}
\end{equation}
In the following, we use the notation
\begin{equation}
D({\imb k}_\perp)\equiv C({\imb k}_\perp)-(2\pi)^2\delta({\imb
k}_\perp)\; .
\end{equation}

To summarize, we have so far obtained a compact expression for the
average of $M_1^{[\widetilde{\rho}]}$ that reads
\begin{eqnarray}
&&\left<M_1^{[\widetilde{\rho}]}({\imb p},{\imb q})\right>_{\widetilde{\rho}}
=i\int {{d^2{\imb k_1}_\perp}\over{(2\pi)^2}}
{{d^2{\imb k_2}_\perp}\over{(2\pi)^2}}
F^{^{QED}}({\imb p}_\perp+{\imb q}_\perp-{\imb k_1}_\perp-{\imb k_2}_\perp)
\nonumber\\
&&\qquad\qquad\qquad\qquad\times
\widetilde{\cal P}({\imb k_1}_\perp+{\imb k_2}_\perp) D({\imb k_2}_\perp)
M({\imb p},{\imb q}|{\imb k_1}_\perp,{\imb k_2}_\perp)\; .
\end{eqnarray}
Note that the factor $\widetilde{\cal P}({\imb k_1}_\perp+{\imb
k_2}_\perp)$ implies that the total momentum transfer ${\imb
k_1}_\perp+{\imb k_2}_\perp$ from the nucleus acting via its color
field is at most of order $1/R$, a feature that was to be expected for
diffraction.

From here, we can proceed as in \cite{GelisP1}. Since the
photoproduction cross-section is enhanced by a factor $Z^2$ for
photons that are produced coherently by the first nucleus, i.e.\ for
photons with a transverse momentum ${\imb p}_\perp+{\imb
q}_\perp-{\imb k_1}_\perp-{\imb k_2}_\perp$ smaller than $1/R$, we can
perform a Taylor expansion of the factor $M({\imb p},{\imb q}|{\imb
k_1}_\perp,{\imb k_2}_\perp)$ around the point where ${\imb
k_1}_\perp+{\imb k_2}_\perp={\imb p}_\perp+{\imb q}_\perp$. Up to
terms of higher order, this gives
\begin{equation}
M({\imb p},{\imb q}|{\imb
k_1}_\perp,{\imb k_2}_\perp)\approx {{\overline{u}({\imb q}) \slv_+ v({\imb p})}\over{p^-+q^-}}+({\imb p}_\perp+{\imb
q}_\perp-{\imb k_1}_\perp-{\imb k_2}_\perp)\cdot {\imb L}({\imb p},{\imb q}|{\imb k_2}_\perp)\; ,
\label{eq: M lin}
\end{equation}
with
\begin{eqnarray}
&&{\imb L}({\imb p},{\imb q}|{\imb k_2}_\perp)\equiv
{{1\over{(p^-+q^-)(m^2+({\imb k_2}_\perp-{\imb p}_\perp)^2)}}}\nonumber\\
&&\qquad\times\overline{u}({\imb q})\Big\{
{{2p^-}\over{p^-+q^-}}
\slv_+({\imb k_2}_\perp-{\imb p}_\perp)+
{{\slv_+ {\imb \gamma}_\perp \slv_- ({\imb k_2}_\perp-{\imb p}_\perp+m)\slv_+}\over{2}}
\Big\}v({\imb p})\; .\nonumber\\
&&
\end{eqnarray}
Thanks to the sum rule\footnote{In order to prove this sum rule, one
has to notice that
\begin{equation}
\int {{d^2{\imb k}_\perp}\over{(2\pi)^2}}\; C({\imb k}_\perp)=
\left<U(0)U^\dagger(0)\right>_{\widetilde{\rho}}=1=\int {{d^2{\imb k}_\perp}\over{(2\pi)^2}}\; (2\pi)^2\delta({\imb k_\perp})\; .
\end{equation}}
\begin{equation}
\int {{d^2{\imb k}_\perp}\over{(2\pi)^2}}\; D({\imb k}_\perp)=0\; ,
\end{equation}
the term of order zero in the expansion (\ref{eq: M lin}) does not
contribute. Integrating over the photon momentum ${\imb
  m}_\perp\equiv{\imb p}_\perp+{\imb q}_\perp-{\imb k_1}_\perp-{\imb
  k_2}_\perp$, while keeping ${\imb k_2}_\perp$ fixed, leads to
\begin{equation}
\left<M_1^{[\widetilde{\rho}]}({\imb p},{\imb q})\right>_{\widetilde{\rho}}=2iZ\alpha e_q \widetilde{\cal P}({\imb p}_\perp+{\imb q}_\perp)\int
{{d^2{\imb k}_\perp}\over{(2\pi)^2}} D({\imb k}_\perp)
{{{\imb b}\cdot{\imb L}({\imb p},{\imb q}|{\imb k}_\perp)}\over{{\imb b}^2}}
\; .
\label{eq:amplitude}
\end{equation}
Using then
\begin{equation}
\int d^2{\imb b}
{{{\imb b}\cdot{\imb L}({\imb p},{\imb q}|{\imb k}_\perp)}\over{{\imb b}^2}}
{{{\imb b}\cdot{\imb L}^*({\imb p},{\imb q}|{\imb k}^\prime_\perp)}\over{{\imb b}^2}}=
{{{\imb L}({\imb p},{\imb q}|{\imb k}_\perp)
\cdot{\imb L}^*({\imb p},{\imb q}|{\imb k}^\prime_\perp)}\over{2}}
\int {{d^2{\imb b}}\over{{\imb b}^2}}
\end{equation}
and
\begin{equation}
\widetilde{\cal P}^2({\imb p}_\perp+{\imb q}_\perp)\approx \pi R^2 (2\pi)^2\delta({\imb p}_\perp+{\imb q}_\perp)\; ,
\end{equation}
we can write the diffractive cross-section for the photoproduction of
a $q\bar{q}$ pair as
\begin{eqnarray}
\sigma_1^{\rm diff}&\equiv&\int d^2{\imb b} P_1^{\rm diff}=\pi R^2 2N_c (Z\alpha)^2 e_q^2 \int\limits_{2R}^{\gamma/m}{{d^2{\imb b}}\over{{\imb b}^2}}\nonumber\\
&&\times\int {{d^3{\imb p}}\over{(2\pi)^3 2\omega_{\imb p}}}
{{d^3{\imb q}}\over{(2\pi)^3 2\omega_{\imb q}}}
(2\pi)^2\delta({\imb p}_\perp+{\imb q}_\perp)
\nonumber\\
&&\times
\int {{d^2{\imb k}_\perp}\over{(2\pi)^2}}
{{d^2{\imb k}^\prime_\perp}\over{(2\pi)^2}}
D({\imb k}_\perp)D({\imb k}^\prime_\perp)
{\rm Tr}\left({\imb L}({\imb p},{\imb q}|{\imb k}_\perp)
\cdot{\imb L}^*({\imb p},{\imb q}|{\imb k}^\prime_\perp)\right) .
\end{eqnarray}
In the integration over the impact parameter, the lower bound is
determined by the geometry of a peripheral collision of two
nuclei of radius $R$, and the upper bound is the limit beyond which
the electromagnetic field does not have a Weizs\"acker-Williams form.
The calculation of the Dirac's trace leads to
\begin{eqnarray}
&&{\rm Tr}\left({\imb L}({\imb p},{\imb q}|{\imb k}_\perp)
\cdot{\imb L}^*({\imb p},{\imb q}|{\imb k}^\prime_\perp)\right)\nonumber\\
&&\qquad=
{{16 p^- q^-}\over{(p^-+q^-)^2 (m^2+({\imb k}_\perp-{\imb p}_\perp)^2)
(m^2+({\imb k}^\prime_\perp-{\imb p}_\perp)^2)}}\nonumber\\
&&\qquad\quad\times\Big\{
m^2+({\imb k}_\perp-{\imb p}_\perp)\cdot({\imb k}^\prime_\perp-{\imb p}_\perp)
\Big[1-2{{p^-}\over{p^-+q^-}}+2{{p^-{}^2}\over{(p^-+q^-)^2}}\Big]
\Big\}\; .\nonumber\\
&&
\end{eqnarray}
At this stage, it is straightforward to perform the integrals over
$p^\pm,q^-$ and ${\imb q}_\perp$ in order to obtain the cross-section
per unit of rapidity (we make use of $dq^+/q^+=2dy$):
\begin{eqnarray}
&&{{d\sigma_1^{\rm diff}}\over{dy}}=\pi R^2 {{4 N_c (Z\alpha)^2 e_q^2}\over{\pi^2}} \int\limits_{2R}^{\gamma/m}{{d^2{\imb b}}\over{{\imb b}^2}}\int  {{d^2{\imb p}_\perp}\over{(2\pi)^2}}
\nonumber\\
&&\times
\int {{d^2{\imb k}_\perp}\over{(2\pi)^2}}
{{d^2{\imb k}^\prime_\perp}\over{(2\pi)^2}}
D({\imb k}_\perp)D({\imb k}^\prime_\perp)
{{m^2+{2\over 3}({\imb k}_\perp-{\imb p}_\perp)\cdot({\imb k}^\prime_\perp-{\imb p}_\perp)}
\over{(m^2+({\imb k}_\perp-{\imb p}_\perp)^2)
(m^2+({\imb k}^\prime_\perp-{\imb p}_\perp)^2)}}\; .\nonumber\\
&&
\end{eqnarray}
One can then note that the function $D({\imb k}_\perp)$ depends only
on the radial variable $k_\perp$, and perform analytically all the
angular integrals, which leads to the following formula for the
$p_\perp$ spectrum of the quarks (or, equivalently, antiquarks)
produced diffractively:
\begin{eqnarray}
&&{{d\sigma_1^{\rm diff}}\over{dydp_\perp}}=\pi R^2 {{N_c (Z\alpha)^2 e_q^2}\over{2\pi^5}} \int\limits_{2R}^{\gamma/m}{{d^2{\imb b}}\over{{\imb b}^2}}\nonumber\\
&&\qquad\qquad\times p_\perp\,
\left[
m^2 I_1^2+{1\over{6p_\perp^2}}(I_2-2p_\perp^2 I_1)^2
\right]\; ,
\label{eq:dsigma_dydpt}
\end{eqnarray}
where we denote\footnote{Note that after integrating over the angle,
  $D$ as a function of the radial variable $k_\perp$ reads:
\begin{equation}
D(k_\perp)=C(k_\perp)-{{4\pi}\over{k_\perp}} \delta(k_\perp)\; .
\end{equation}}
\begin{eqnarray}
&&I_1\equiv\int\limits_0^{+\infty} dk_\perp k_\perp D(k_\perp)
{1\over{\sqrt{(m^2+p_\perp^2+k_\perp^2)^2-4p_\perp^2 k_\perp^2}}}\; ,\nonumber\\
&&I_2\equiv\int\limits_0^{+\infty} dk_\perp k_\perp D(k_\perp)
{{m^2+p_\perp^2+k_\perp^2}\over{\sqrt{(m^2+p_\perp^2+k_\perp^2)^2-4p_\perp^2 k_\perp^2}}}\; .
\end{eqnarray}
\begin{figure}[htb]
\centerline{
\resizebox*{!}{6.5cm}{\includegraphics{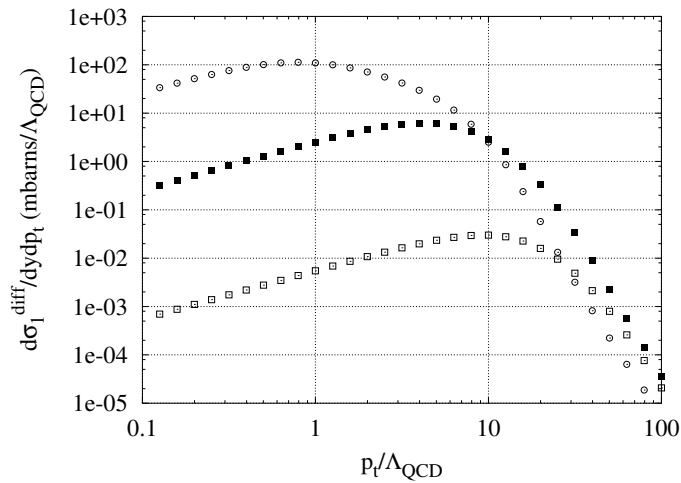}}}
\caption{\label{fig:pt-spectrum} Differential diffractive
  cross-section as a function of the transverse momentum of the quark.
  The value of $Q_s/\Lambda_{_{QCD}}$ is set to $10$. Open circles:
  strange quarks. Filled squares: charm quark. Open squares: bottom
  quark. The differential cross-section is in units of
  mbarns$/\Lambda_{_{QCD}}$.}
\end{figure}
Those two integrals are easy to evaluate numerically in order to
obtain the $p_\perp$ spectrum of the produced quarks.  Results are
plotted in figure \ref{fig:pt-spectrum} for a value of
$Q_s/\Lambda_{_{QCD}}=10$ (i.e.\ $Q_s\approx 2$GeV, if we assume
$\Lambda_{_{QCD}}\approx 200$MeV) for strange
($m/\Lambda_{_{QCD}}\approx 1$), charm ($m/\Lambda_{_{QCD}}\approx 8$)
and bottom ($m/\Lambda_{_{QCD}}\approx 23$) quarks. Corresponding to
LHC collision energy, we chose a Lorentz factor of $\gamma=3000$,
which is in the regime where the leading logarithm approximation is
valid.

\section{Integrated diffractive cross-section}
\label{sec:cross-section}
\subsection{Cross-section}
\begin{figure}[htb]
\centerline{
\resizebox*{!}{6.5cm}{\includegraphics{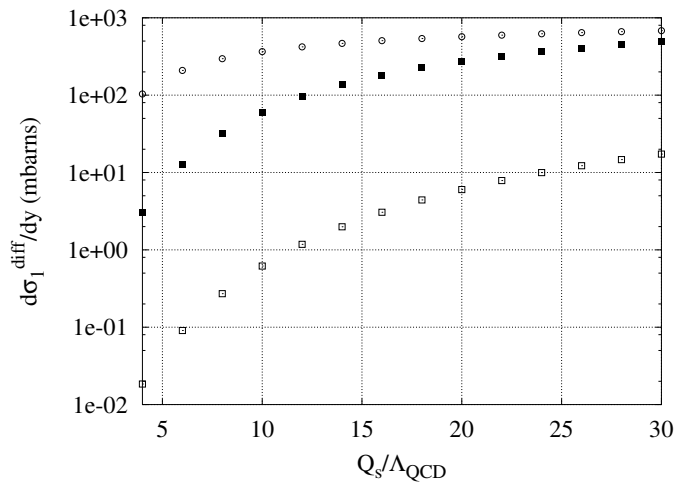}}}
\caption{\label{fig:dsdy} Diffractive cross-section in mbarns per unit 
  of rapidity as a function of the saturation scale. Open circles: 
  strange quarks. Filled squares: charm quarks. Open squares: bottom 
  quarks.}
\end{figure}

The next step is to perform the integration over the transverse
momentum of the produced quarks, in order to obtain total
cross-sections or production probabilities. This can be done
numerically, and the results are plotted in figure \ref{fig:dsdy} as a
function of $Q_s/\Lambda_{_{QCD}}$, for strange, charm, and bottom
quarks. Again, the Lorentz factor is taken to be $\gamma=3000$.
We can see that the cross-section for the production of $c\bar{c}$ and
$b\bar{b}$ pairs is very sensitive to the value of the saturation
scale $Q_s$, especially in the range $Q_s/\Lambda_{_{QCD}}\in[5,15]$.
Therefore, this process could be used as a way to measure the
saturation scale for ultra-relativistic heavy nuclei.

\subsection{Unitarity and black-disc limit}
\begin{figure}[ht]
\centerline{
\resizebox*{!}{5cm}{\includegraphics{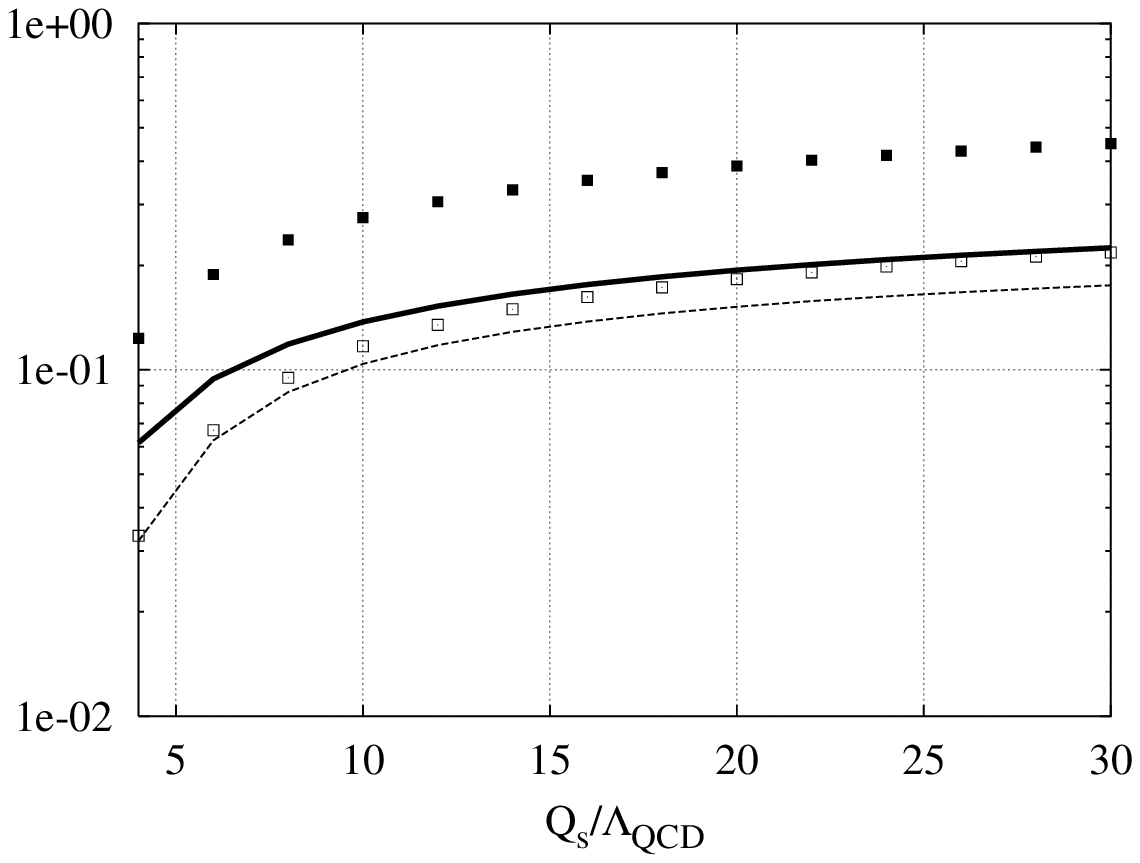}}}
\centerline{
\resizebox*{!}{5cm}{\includegraphics{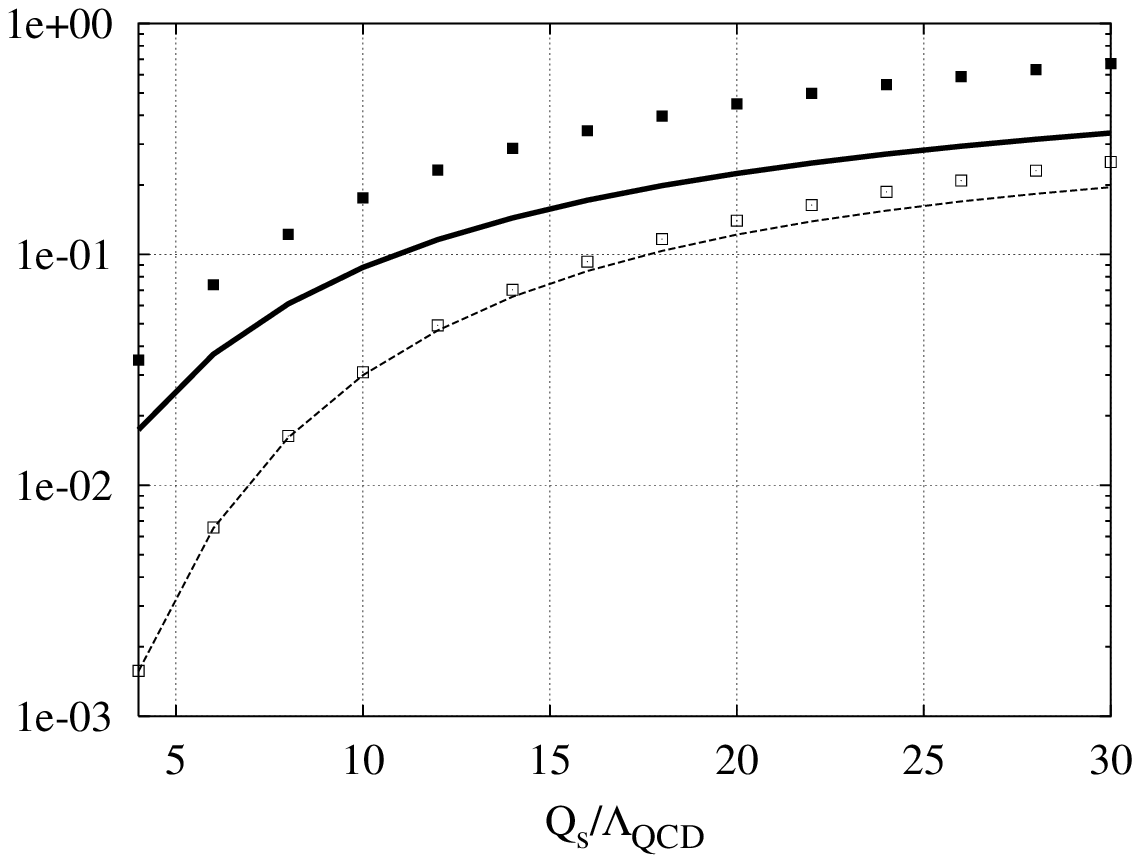}}}
\caption{\label{fig:upper-limit-check} Verification of
  the inequality $P_1^{\rm diff}\le \overline{n}/2$ for the production
  of $q\bar{q}$ pairs at $b=2R$, as a function of $Q_s/\Lambda_{_{QCD}}$.
  Top: strange quarks; bottom: charm quarks. Filled squares: average
  multiplicity for inclusive events ($\overline{n}$); solid line:
  $\overline{n}/2$. Open squares: non-unitarized probability for the
  production of a $q\bar{q}$ pair in a diffractive event ($P_1^{\rm
    diff}$); dashed line: naive unitarization of this probability
  ($P_1^{\rm diff}\exp(-P_1^{\rm diff})$).}
\end{figure}

It is interesting to investigate explicitly the inequality
(\ref{eq:inequal}) between the diffractive $q\bar{q}$ production
probability $P_1^{\rm diff}$ and the inclusive probability $P_1^{\rm
  incl}$ (which at leading order in $Z\alpha$ is equal to
$\overline{n}$ calculated in \cite{GelisP1}). The probability can
easily be obtained from the expression for the respective
cross-section by integrating over the rapidity, but without performing
the integration over the impact parameter. Note that assuming $\gamma
\rightarrow \infty$ in our present approach leads to a flat rapidity
distribution, and a cutoff in rapidity must be introduced by hand. For
the numerical calculation, we assume\footnote{The precise value of the
  cutoff does not play any role for the verification of the inequality
  between $P_1^{\rm diff}$ and $P_1^{\rm incl}/2$, it has only an
  influence on the size of the unitarization corrections.} a rapidity
range of $2\ln(\gamma)$, which leads to the relation
\begin{equation}
 P_1^{\rm diff}(2R)
 =
 \ln\gamma \left[ 4\pi R^2 \ln\frac\gamma{2mR} \right]^{-1}
 \frac{d\sigma_1^{\rm diff}}{dy}
\end{equation}
between the probability at the smallest possible impact parameter and
the cross-section. In figure \ref{fig:upper-limit-check}, we have
displayed $\overline{n}$, $\overline{n}/2$ and $P_1^{\rm diff}$, at
$b=2R$ and for a Lorentz factor $\gamma=3000$, as functions of
$Q_s/\Lambda_{_{QCD}}$ for strange and charm quarks.
As expected, we always find $P_1^{\rm diff}\le \overline{n}/2$. We
also observe that the two quantities tend to a common value as $Q_s$
becomes large. Indeed, in the McLerran-Venugopalan model the
black-disc limit corresponds to an infinitely large value of the
saturation scale.

We have already emphasized the importance of the prefactor $\left<
  0_{\rm out}|0_{\rm in} \right>_{\widetilde{\rho}}$, to ensure that
the obtained probabilities are smaller than $1$. However, in the
present calculation, in which we have kept only the terms of lowest
order in $Z\alpha$, this factor is approximated by $1$. Therefore, it
is expected that this approximation violates unitarity, and that
unitarity is restored by terms of higher order in the electromagnetic
coupling.  These corrections are rather difficult to calculate
exactly.  For a rough estimate of their size, we suppose that the
diffractive multiplicities have a Poissonian
distribution\footnote{This supposition does not hold strictly due to
  correlations in the amplitude to produce more than one pair, see
  \cite{BaltzGMP1}.}, and that the unitarized probability to create
one pair diffractively is given by $P_1^{\rm diff}\exp(-P_1^{\rm
  diff})$.  This expression is also plotted in figure
\ref{fig:upper-limit-check}.  We see that the unitarization effects
(i.e.\ the higher order corrections in $Z\alpha$) are not important as
long as the average multiplicity is much smaller than one. In
particular, we estimate that the unitarity corrections remain small in
the range of values of $Q_s$ expected to be relevant at RHIC or at LHC
(between $1$ and $3$GeV). We also emphasize the fact that this
estimate has been done for the probabilities at the smallest allowed
impact parameter $b=2R$, where the non-unitarized probabilities are
the largest. Integrating the unitarized probability over the impact
parameter for an estimate of the unitarization corrections in the
diffractive cross-section, they are found to be smaller by a relative
factor of $P_1^{\rm diff}(2R) / 2 \ln(\gamma/2mR)$. Hence, they can be
neglected here.

\subsection{Summary}
To close this section, let us provide a table of the values of the
cross-section per unit of rapidity, for charm and bottom quark pairs,
and for inclusive\footnote{The inclusive values come from
\cite{GelisP1}, corrected by the appropriate factors of $e_q^2$ for
the electric charges of the quarks, which were not included in
\cite{GelisP1}.} and diffractive production:
\begin{center}
\begin{tabular}{|c|c|c|}
\hline
& charm & bottom \\
\hline
$d\sigma_{\rm incl}/dy\ (\rm mb) $ & 355 & 11\\
\hline
$d\sigma_{\rm diff}/dy\ (\rm mb) $ &  60 & 0.62\\
\hline
\end{tabular}
\end{center}
Those numbers are obtained for a Lorentz factor $\gamma=3000$, a
nuclear radius $R=6$fm and a saturation scale
$Q_s=10\Lambda_{_{QCD}}$.

\section{Conclusions}
\label{sec:conclusion}
In this paper, we have evaluated the cross-section for the diffractive
photoproduction of $q\bar{q}$ pairs in peripheral collisions of heavy
nuclei. Diffractive events are particularly interesting because they
could provide a cleaner final state due a rapidity gap between the
nuclei and the pair.  This process can be used as a way to test
saturation models, and to measure the value of the saturation scale in
heavy ion experiments. Indeed, the yield of $q\bar{q}$ pairs has been
found to be very sensitive to the value of the saturation scale,
especially for $c$ and $b$ quarks and for values of the saturation
scale between $1$ and $3$GeV.

We have also checked that our result is consistent with the one
obtained in \cite{GelisP1} for the photoproduction of pairs in
inclusive events. In particular, the inequality $P_1^{\rm diff}\le
P_1^{\rm incl}/2$ is satisfied by our calculation, and the black-disc
limit is recovered when the saturation scale becomes infinite.

\noindent{\bf Acknowledgments:} This work is supported by DOE under
grant DE-AC02-98CH10886.  A.P. is supported in part by the
A.-v.-Humboldt foundation (Feo\-dor-Lynen program). We would like to
thank A. Dumitru, J. Jalilian-Marian, D. Kharzeev, L. McLerran, R.
Venugopalan and U. Wiedemann for discussions on this problem.

\appendix

\bibliographystyle{unsrt} 


\begin{thebibliography}{10}

\bibitem{Muell4}
{A.H. Mueller}, Lectures given at the International Summer School on Particle
  Production Spanning MeV and TeV Energies (Nijmegen 99), Nijmegen,
  Netherlands, 8-20, Aug 1999, hep-ph/9911289.

\bibitem{McLer1}
{L. McLerran}, Lectures given at the 40'th Schladming Winter School: Dense
  Matter, March 3-10 2001, hep-ph/0104285.

\bibitem{GriboLR1}
{L.V. Gribov, E.M. Levin, M.G. Ryskin}, Phys. Rept. {\bf 100}, 1 ({1983}).

\bibitem{MuellQ1}
{A.H. Mueller, J-W. Qiu}, Nucl. Phys. {\bf B} {\bf 268}, 427 ({1986}).

\bibitem{FrankS1}
{L.L. Frankfurt, M.I. Strikman}, Phys. Rept. {\bf 160}, 235 ({1988}).

\bibitem{Lipat1}
{L.N. Lipatov}, Sov. J. Nucl. Phys. {\bf 23}, 338 ({1976}).

\bibitem{KuraeLF1}
{E.A. Kuraev, L.N. Lipatov, V.S. Fadin}, Sov. Phys. JETP {\bf 45}, 199
  ({1977}).

\bibitem{BalitL1}
{I. Balitsky, L.N. Lipatov}, Sov. J. Nucl. Phys. {\bf 28}, 822 ({1978}).

\bibitem{JalilKMW1}
{J. Jalilian-Marian, A. Kovner, L. McLerran, H. Weigert}, Phys. Rev. {\bf D}
  {\bf 55}, 5414 ({1997}).

\bibitem{KovchM1}
{Yu.V. Kovchegov, L. McLerran}, Phys. Rev. {\bf D} {\bf 60}, 054025 ({1999}).

\bibitem{KovchM2}
{Yu.V. Kovchegov, L. McLerran}, Erratum Phys. Rev. {\bf D} {\bf 62}, 019901
  ({2000}).

\bibitem{McLerV1}
{L. McLerran, R. Venugopalan}, Phys. Rev. {\bf D} {\bf 49}, 2233 ({1994}).

\bibitem{McLerV2}
{L. McLerran, R. Venugopalan}, Phys. Rev. {\bf D} {\bf 49}, 3352 ({1994}).

\bibitem{McLerV3}
{L. McLerran, R. Venugopalan}, Phys. Rev. {\bf D} {\bf 50}, 2225 ({1994}).

\bibitem{JalilKLW1}
{J. Jalilian-Marian, A. Kovner, A. Leonidov, H. Weigert}, Nucl. Phys. {\bf B}
  {\bf 504}, 415 ({1997}).

\bibitem{JalilKLW2}
{J. Jalilian-Marian, A. Kovner, A. Leonidov, H. Weigert}, Phys. Rev. {\bf D}
  {\bf 59}, 014014 ({1999}).

\bibitem{JalilKLW3}
{J. Jalilian-Marian, A. Kovner, A. Leonidov, H. Weigert}, Phys. Rev. {\bf D}
  {\bf 59}, 034007 ({1999}).

\bibitem{JalilKLW4}
{J. Jalilian-Marian, A. Kovner, A. Leonidov, H. Weigert}, Erratum. Phys. Rev.
  {\bf D} {\bf 59}, 099903 ({1999}).

\bibitem{KovneM1}
{A. Kovner, G. Milhano}, Phys. Rev. {\bf D} {\bf 61}, 014012 ({2000}).

\bibitem{KovneMW3}
{A. Kovner, G. Milhano, H. Weigert}, Phys. Rev. {\bf D} {\bf 62}, 114005
  ({2000}).

\bibitem{Balit1}
{I. Balitsky}, Nucl. Phys. {\bf B} {\bf 463}, 99 ({1996}).

\bibitem{Kovch3}
{Yu.V. Kovchegov}, Phys. Rev. {\bf D} {\bf 61}, 074018 ({2000}).

\bibitem{IancuLM1}
{E. Iancu, A. Leonidov, L. McLerran}, Nucl. Phys. {\bf A} {\bf 692}, 583
(2001).

\bibitem{IancuLM2}
{E. Iancu, A. Leonidov, L. McLerran}, Phys. Lett. {\bf B} {\bf 510}, 133
  ({2001}).

\bibitem{Muell5}
{A.H. Mueller}, Phys. Lett. {\bf B} {\bf 523}, 243 (2001).

\bibitem{GolecW1}
{K. Golec-Biernat, M. W\"usthoff}, Phys. Rev. {\bf D} {\bf 59}, 014017 ({1999}).

\bibitem{GolecW2}
{K. Golec-Biernat, M. W\"usthoff}, Phys. Rev. {\bf D} {\bf 60}, 114023 ({1999}).

\bibitem{GolecW3}
{K. Golec-Biernat, M. W\"usthoff}, Eur. Phys. J. {\bf C} {\bf 20}, 313 ({2001}).

\bibitem{GotsmLLMT1}
{E. Gotsman, E. Levin, M. Lublinsky, U. Maor, K. Tuchin}, hep-ph/0007261.

\bibitem{GotsmLLMT2}
{E. Gotsman, E. Levin, M. Lublinsky, U. Maor, K. Tuchin},
Phys. Lett. {\bf B} {\bf 492}, 47  (2000).
 
\bibitem{McLerV4}
{L. McLerran, R. Venugopalan}, Phys. Rev. {\bf D} {\bf 59}, 094002 ({1999}).

\bibitem{KovneMW1}
{A. Kovner, L. McLerran, H. Weigert}, Phys. Rev. {\bf D} {\bf 52}, 3809
  ({1995}).

\bibitem{KovneMW2}
{A. Kovner, L. McLerran, H. Weigert}, Phys. Rev. {\bf D} {\bf 52}, 6231
  ({1995}).

\bibitem{Kovch4}
{Yu.V. Kovchegov}, Nucl. Phys. {\bf A} {\bf 692}, 557 (2001).

\bibitem{KrasnV1}
{A. Krasnitz, R. Venugopalan}, Phys. Rev. Lett. {\bf 84}, 4309 ({2000}).

\bibitem{KrasnV2}
{A. Krasnitz, R. Venugopalan}, Phys. Rev. Lett. {\bf 86}, 1717 ({2001}).

\bibitem{GelisP1}
{F. Gelis, A. Peshier}, Nucl. Phys. {\bf A} {\bf 697}, 879 (2002).

\bibitem{KovchL1}
{Yu.V. Kovchegov,E. Levin}, Nucl. Phys. {\bf B} {\bf 577}, 221 ({2000}).

\bibitem{BaltzGMP1}
{A.J. Baltz, F. Gelis, L. McLerran, A. Peshier}, Nucl. Phys. {\bf A} {\bf 695},
  395 ({2001}).

\bibitem{BuchmGH1}
{W. Buchm\"uller, T. Gehrmann, A. Hebecker}, Nucl. Phys. {\bf B} {\bf 537}, 477
  ({1999}).

\bibitem{BuchmMH1}
{W. Buchm\"uller, M.F. McDermott, A. Hebecker}, Phys. Lett. {\bf B} {\bf 410},
  304 ({1997}).

\bibitem{KovneW1}
{A. Kovner, U. Wiedemann}, Phys. Rev. {\bf D} {\bf 64}, 114002 ({2001}).

\bibitem{LamM1}
{C.S. Lam, G. Mahlon}, Phys. Rev. {\bf D} {\bf 62}, 114023 ({2000}).

\end{thebibliography}

\end{document}